\RequirePackage[2020-02-02]{latexrelease}
\documentclass[preprintnumbers,amsmath, amssymb, superscriptaddress, pra]{revtex4}
\usepackage{graphicx}
\usepackage{subfigure}
\usepackage{dcolumn}
\usepackage{bm}
\usepackage{epsf}
\usepackage{amssymb}
\DeclareGraphicsExtensions{.eps}
\begin{document}

\author{David S. Simon}
\email[e-mail: ]{simond@bu.edu} \affiliation{Dept. of Physics and Astronomy, Stonehill College, 320 Washington Street, Easton, MA 02357} \affiliation{Dept. of
Electrical and Computer Engineering \& Photonics Center, Boston University, 8 Saint Mary's St., Boston, MA 02215, USA}
\author{Christopher R. Schwarze}
\email[e-mail: ]{crs2@bu.edu} \affiliation{Dept. of Electrical and Computer Engineering \& Photonics Center, Boston University, 8 Saint Mary's St., Boston, MA
02215, USA}
\author{Abdoulaye Ndao}\email[e-mail: ]{a1ndao@ucsd.edu} \affiliation{Dept. of Electrical and Computer Engineering \& Photonics Center, Boston University, 8 Saint Mary's St., Boston, MA
02215, USA}\affiliation{Department of Electrical and Computer Engineering, University of California San Diego, La Jolla, CA, USA}
\author{Alexander V. Sergienko}
\email[e-mail: ]{alexserg@bu.edu} \affiliation{Dept. of Electrical and Computer Engineering \& Photonics Center, Boston University, 8 Saint Mary's St., Boston,
MA 02215, USA} \affiliation{Dept. of Physics, Boston University, 590 Commonwealth Ave., Boston, MA 02215, USA}




\begin{abstract}
The Su-Schrieffer-Heeger (SSH) system is a popular model for exploring topological insulators and topological phases in one dimension. Recent interest in exceptional points has led to re-examination of non-Hermitian generalizations of many physical models, including the SSH model. In such non-Hermitian systems, singular points called exceptional points (EPs) appear that are of interest for applications in super-resolution sensing systems and topological lasers. Here, a non-Hermitian and non-${\cal P}{\cal T}$-symmetric variation of the SSH model is introduced, in which the hopping amplitudes are non-reciprocal and vary monotonically along the chain. It is found that, while the existence of the EPs is due to the nonreciprocal couplings, the number, position, and order of the EPs can all be altered by the addition of the hopping amplitude gradient, adding a new tool for tailoring the spectrum of a non-Hermitian system.

\end{abstract}

\title{Exceptional points in SSH-like models with hopping amplitude gradient}
\maketitle

\section{Introduction}
Traditionally, research in quantum mechanics has focused on Hermitian systems, since they come with real eigenvalues that can serve as physical observables. But the discovery \cite{bender1,bender2,bender3} that the weaker condition of parity-time (${\cal P}{\cal T}$) invariance is sufficient to ensure real eigenvalues has caused researchers to re-examine the physics of non-Hermitian Hamiltonians \cite{oz,feng}. Eigenvalue degeneracies in Hermitian systems typically correspond to singularities called diabolic points (DPs). At the diabolic point, two or more of the eigenvalues become equal, while the dimension of the space of eigenvectors remains unchanged. As the system is perturbed away from the DP, the splitting between the energy levels is proportional to the perturbation strength $\epsilon$, for small values of  $\epsilon$. A simple example of a DP occurs in the Bohr model of hydrogen. In the absence of any external fields, the 2s and 2p orbitals of the atom are degenerate. However when an electric field is turned on, the new eigenstates become 2s-2p hybrid states, and the energy levels split. The electric field magnitude plays the role of the perturbation parameter, and for small fields the energy splittings increase proportional to $\epsilon $. The $\epsilon =0$ point is the diabolical point: at this point the energies are degenerate, but the eigenstates remain distinct.

Non-Hermitian systems, having inherently complex spectra, allow for the possibility of an additional type of singularity, called an exceptional point (EP) or non-Hermitian degeneracy \cite{kato,berry1,berry2,heiss1,heiss2,heiss3}, that does not appear in Hermitian quantum systems. As the EP is approached in parameter space, not only do two or more eigenvalues degenerate, but the eigenvectors coalesce as well, causing a drop in the dimension of the eigenspace. Generally, this is accompanied by a transition from real to complex eigenvalues. Initially, these EPs were viewed as a nuisance, since they form obstacles that limit the range of validity of perturbative expansions. But, more recently, they have come to be of great interest due to the rapid variation in eigenvalue splitting when perturbed away from the EP by a small amount \cite{hodaei,hokma,park,zhao,miri}. This strong response to very small perturbations makes them useful for high-sensitivity sensors \cite{hodaei,wier1,wier3,wier4,liu2,chen}. Exceptional point sensors have now been developed in a wide variety of physical platforms. Examples include microcavity particle detectors \cite{voll}, magnetic field sensors \cite{rondin}, and optical gyroscopes \cite{chow,sunada}, and other photonic sensing systems \cite{ganainy,oz,miri,wier1}. In addition, ${\cal P}{\cal T}$ symmetry and its breaking have been used to develop single-mode microring lasers \cite{feng1,hodaei1,miao} and to generate optical orbital angular momentum (OAM) states \cite{lin,alex}.

The number of eigenvectors that coalesce is called the \emph{order} of the exceptional point \cite{heiss3}. The simplest case is a second-order EP, which has an energy level splitting that scales $\sim \epsilon^{1/2}$ as a function of the perturbation strength $\epsilon$; the low exponent implies that the levels vary faster than any polynomial-order perturbation for small $\epsilon$.  More generally, an $n$th-order EP has scaling that is enhanced scale to give an overall sensitivity of $\epsilon^{1/ n}$.

Although there has been some debate over whether the improved sensitivity leads to real sensing improvement once the similarly-enhanced noise is taken into account \cite{langbein,lau,chen2,zhang,wang,smith,weir2,konon}, evidence (based on increased Fisher information arguments) seems to be that the improvement remains significant \cite{anderson} under some circumstances. In any case, improved understanding of EPs and the ability to produce them in different physical platforms continue to be topics of great interest.

Periodic modulations and their effects on topological properties in both Hermitian \cite{bukov,holt,eck,wein,oka} and non-Hermitian \cite{zhou} Floquet systems have been well-studied, but the effect of monotonic parameter changes has  been less examined. Here it is shown that even such monotonic variations can lead to useful effects related to the topological and singularity structure of the system. This will be demonstrated using a non-Hermitian variation on the one-dimensional Su-Schreiffer-Heeger (SSH) system, in which it will be shown that allowing a nonzero gradient can alter the singularity structure, and in particular, it can change number and position of EPs and can cause two lower-order EPs to merge into a higher-order EP.

The linear parameter gradient is a fairly general situation in the sense that if the hoppings vary according to any analytic function of position, they can always be approximated for slow spatial variation by the first few terms of a Taylor expansion. The linear term will be the leading non-trivial contribution, and perturbations to this term will control the overall behavior of the system near singular points. So, the behavior seen here should be generic for slow parameter variations.

In the next section, we give a brief review of the SSH model and some of its non-Hermitian generalizations. Then in Section 3 we introduce the specific model studied in this paper, in which both non-reciprocal couplings and a variation of the couplings are introduced along the chain.  Our main result is that the introduction of the coupling gradient can alter the EP structure, causing EPs to collide or bifurcate, as well as changing their locations in parameter space. In particular, the collision of two second order EPs can result in the creation of a fourth order EP. We briefly potential experimental implementations in optical and electronic circuit platforms in Section 4, before discussing some conclusions in Section 5.

\section{The SSH model and its extensions}\label{SSHsection}

The standard Hermitian Su-Schrieffer-Heeger (SSH) model \cite{ssh,asboth,batra} is widely used as a simple example of a one-dimensional topological insulator. It was originally proposed as a model of the polyacetylene molecule \cite{ssh}, but has since been realized in physical platforms ranging from cold atoms and photonic systems to electric circuits
\cite{atala,karski,broome,schreiber, kitagawa,liu}. SSH models have gained increased importance since lasing has been demonstrated in a one-dimensional lattice of polariton micropillars that realizes the SSH Hamiltonian \cite{stjean}.
Lasing has since been shown to also exist in other non-Hermitian systems \cite{ge2}.

\begin{figure}
\centering
\includegraphics[totalheight=0.5in]{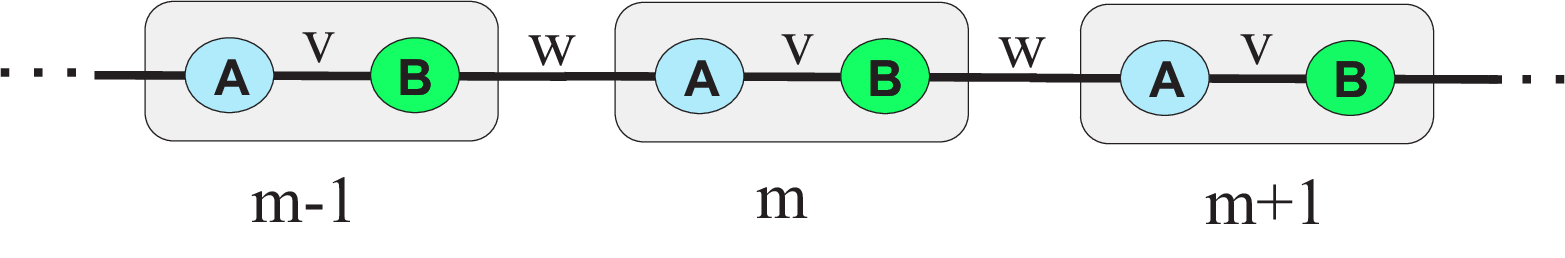}
\caption{The SSH model: each unit cell consists of two distinct subsites or substates, $A$ and $B$. $v$ is the
amplitude for transition leftward from $B$ to $A$ within the same unit cell, while $w$ describes leftward transitions of $A$ in cell  $m$ to $B$ of cell $m-1$. Hermiticity requires that the transitions in the opposite directions along the same bonds to be $v^\ast$ and $w^\ast$. In the non-Hermitian case this constraint will no longer be hold.}\label{sshmodelfig}
\end{figure}

The model discussed here consists of a chain of unit cells, each consisting of two distinct lattice sites. Labelling the unit cells by integer $m$, the two lattice sites within are labelled $A$ and $B$.  Nearest-neighbor hopping amplitudes between the sites alternate, with $v$ being the amplitude to hop between $B$ and $A$ sites within the same cell, and $w$ the amplitude to hop between $A$ of cell $m$ and the $B$ site of the adjacent cell, $m-1$ (Figure~\ref{sshmodelfig}). Hermiticity requires that the hopping amplitudes be reciprocal: the amplitude to hop to the left between a pair of cells should be the complex conjugate of the amplitude to hop to the right between the same pair, so that the SSH Hamiltonian takes the form
\begin{eqnarray}H &=& \sum_m \big( v |A,m\rangle \langle B,m | +w |B,m-1\rangle \langle A,m |  \\ & & \qquad +  v^\ast |B,m\rangle \langle A,m | +w^\ast |A,m\rangle \langle B,m-1 |\big)  .\nonumber  \end{eqnarray}
In the Hermitian case, it is possible, by redefining the phase of the wavefunctions if necessary, to take the amplitudes to be real, $v=v^\ast$ and $w=w^\ast$. Assuming real amplitudes, the SSH model Hamiltonian can be put into matrix form,
\begin{equation}H=\left( \begin{array}{cccccc} 0 & v & 0 & 0 &0 & 0\\ v & 0 & w& 0 &0  &0 \\ 0 & w & 0& v& 0 &  0 \\
0 & 0 & v& 0 & w & 0 \\  0 & 0 & 0 & w & 0 & v \\  0 & 0 & 0 &  0 &  \ddots & \ddots\\
 \end{array}  \right) ,\end{equation} where the odd-numbered rows and columns represent the $A$ sites at consecutive $m$ values, while the even-numbered rows and columns represent the corresponding $B$ sites. (Note that the signs, or more generally the phases, of the couplings can easily be controlled in optical implementations by changing the distance between lattice sites, by inserting phase shifters, or by altering the dispersion of the material between the sites. For simplicity, we keep the couplings real but not necessarily positive.)

The SSH model has a chiral sublattice symmetry, $\Gamma H\Gamma^\dagger =-H$, given by $\Gamma = \hat P_A-\hat P_B$, where $P_{A,B}$ are projectors onto the $A$ and $B$ sublattices. Because of the chiral symmetry, all nonzero eigenvectors come in opposite sign pairs. More importantly, it has implications for the topological properties of the system. Up to an overall additive constant, the momentum space Hamiltonian can be written as $H(k) =\bm d(k)\cdot \bm \sigma$, where $k$ is wavenumber and $\sigma_j$ are the Pauli matrices. The chiral symmetry forces the complex vector $\bm d(k)$ to lie in the $x$-$y$ plane. In order for the material to be an insulator, with a nonzero energy gap, the remaining two-dimensional vector must avoid the origin.  So as $k$ varies across the Brillouin zone, $d(k)$ traces out a closed curve, which circles the origin either once, or not at all. In other words, the system has a winding number of either ${\cal V} =1$ or ${\cal V} =0$ about the origin. The system's properties are unusually stable with respect to perturbations, because transitions from one topological phase to the other cannot occur unless the energy gap closes, which can only happen if the chiral symmetry is broken. It can easily be shown that the ${\cal V}=0$ phase occurs when $|v|<|w|$, while ${\cal V}=1$ occurs when $|v|>|w|$.

In the model studied here, we will for simplicity assume that all the parameters in the Hamiltonian are real. The fact that the Hamiltonian will then be real will ensure the existence of time reversal invariance. The chiral and time reversal invariances will together guarantee that the energy spectrum is symmetric under reflections across both the real and imaginary axes. When the parameters are allowed to be complex, new effects could potentially appear, but this is a topic for later investigation.

For periodic boundary conditions or for infinite-length chains, the choice of which parameter to call $w$ and which to call $v$ is arbitrary. But for finite chains with open boundary conditions (the case considered here), $v$ is always the amplitude on the last link at either end of the chain. The topologically-nontrivial case is then the one in which the hoppings across those end links are weak, encouraging isolation of the lattice sites at the ends of the chain.

Several non-Hermitian variations on the SSH model have been studied \cite{yao,song,he1,zhu2,lieu}. Ref. \cite{he1} looks at non-Hermitian models with longer range hopping (longer than nearest neighbor) and Ref. \cite{zhu2} looks at non-Hermitian, ${\cal P}{\cal T}$-symmetric models with complex boundary potentials. Ref. \cite{lieu} examined models which are non-reciprocal (left-moving and right-moving hoppings are not equal) but not ${\cal P}{\cal T}$-symmetric; these were shown to share several properties with PT-symmetric models. In the non-Hermitian case, the amplitudes for hopping to the left and to the right are not necessarily complex conjugates of each other, \begin{eqnarray}H&=&\sum_m \big( v_L |A,m\rangle \langle B,m | +w_L |B,m-1\rangle \langle A,m |  \\ & & \qquad  +  v_R |B,m\rangle \langle A,m | +w_R |A,m\rangle \langle B,m-1 | \big) ,\nonumber\end{eqnarray} with $w_L\ne w_R^\ast$ and $v_L\ne v_R^\ast$. Non-Hermiticity commonly arises in open systems, where gain and loss can occur.

Here, we look at finite-length SSH-like models with nearest-neighbor hoppings and open boundary conditions at the ends. Two parameters are added to the model: one to control the amount of non-reciprocity in hopping amplitudes, and one to control the rate of variation of hopping amplitudes with position along the chain. We assume the simplest case, with constant hopping gradient, or in other words, an increase or decrease of hopping amplitude that varies linearly along the length of the chain. Exceptional points (EPs) will form continuous families in the parameter space, and for long chains the structure of the spectrum and the distribution of the EPs can become highly complex, with the number and order of the EPs grows as the SSH chain becomes longer. The nonzero gradient breaks the discrete translational symmetry of the system; as has been seen in other systems \cite{cui,kodigala,lourenco,park,abdo}, the breaking of spatial symmetries often leads to the appearance of EPs.

There is a rich landscape of variations of related models. Here we look at a simple model that seems to be fairly representative, focusing on short chain lengths for which analytic expressions can be found for the eigenvectors and spectrum. Because of the lack of translational invariance, it is more convenient to work in position space, rather than in momentum space.

\begin{figure}
\centering
\includegraphics[totalheight=1.2in]{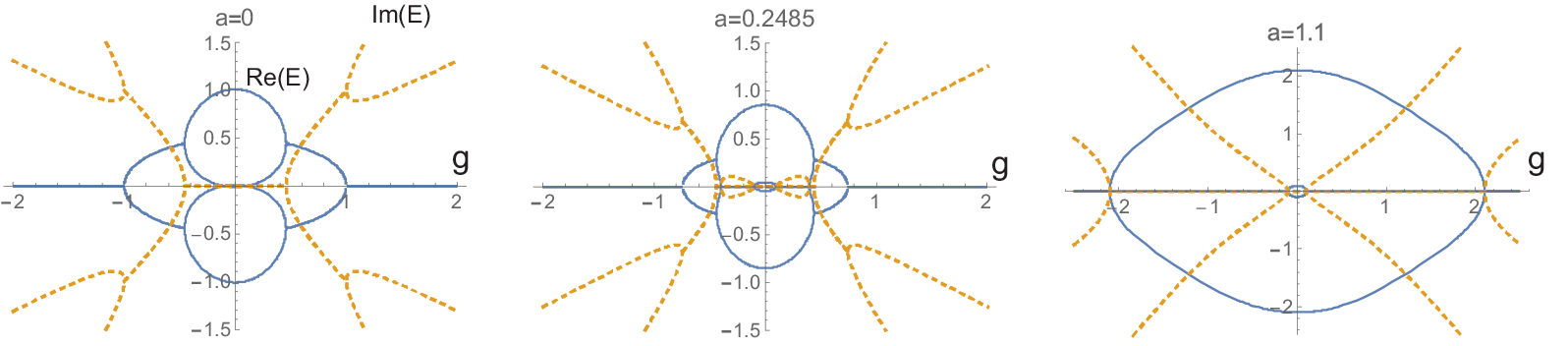}
\caption{Examples of the spectrum for $|v|\le |w|$. Here the values $v=0.1$, $w=1$ were used. Energies are plotted versus non-reciprocity parameter $a$ for several values of gradient parameter $g$. The solid blue curve is the real part of the energy and the dashed yellow curve is the imaginary part. }\label{vlewspecfig}
\end{figure}

\section{The non-reciprocal linear gradient model}

Take a one-dimensional lattice with hoppings between nearest-neighbor sites, and assume a position-space Hamiltonian of the form
\begin{equation}H=\left( \begin{array}{cccccc} 0 & v+g & 0 & 0 &0 & 0\\ v-g & 0 & w+g-a& 0 &0  &0 \\ 0 & w-g-a & 0& v+g-2a& 0 &  0 \\
0 & 0 & v-g-2a& 0 & w+g-3a & 0 \\  0 & 0 & 0 & w-g-3a & 0 & v+g-4a \\  0 & 0 & 0 &  0 &  \ddots & \ddots\\
\end{array}  \right) .\label{HGen}\end{equation}  For $g\ne 0$ the model is non-reciprocal, with different probabilities for hopping to the left and to the right. This model is neither Hermitian nor ${\cal P}{\cal T}$-symmetric for $a\ne 0$, but reduces back to the SSH model for $a=g=0$.    Because of the asymmetry of the hopping coefficients, the amplitude will tend (for open boundary conditions) to collect at one end of the lattice, in accord with the non-Hermitian skin effect \cite{yao,alvarez}. Such boundary states exhibit new effects not seen in Hermitian systems \cite{qi,ge1,riv,hent}. The additional parameter $a$ controls how fast the hopping rates vary along the chain. We will restrict ourselves here to the case where all parameters ($v$, $w$, $a$, and $g$) are real. For $a=0$ and $v=w$, the model examined here reduces to the Hatano-Nelson \cite{hn1,hn2}.
Recently, Ref. \cite{ge0} demonstrated a method of generating non-Hermitian Hamiltonians with real spectra from Hermitian Hamiltonians; this approach can generate both spatially varying hopping coefficients and nonreciprocality. The current paper can be viewed as a case of this more general setup with vanishing on-site potential and a particular choice of hopping coefficient variations. Other non-reciprocal models have been studied on occasion in platforms ranging from photonic and acoustic systems to mechanical systems \cite{gao, longhi, zhang1, wang1}.

For now, focus on the simplest nontrivial case, of four lattice sites, spanning two unit cells:
\begin{equation}H=\left( \begin{array}{cccccc} 0 & v+g & 0 & 0 \\ v-g & 0 & w+g-a& 0 \\ 0 & w-g-a & 0& v+g-2a \\
0 & 0 & v-g-2a& 0 \end{array}  \right) .\label{H4}\end{equation}

\begin{figure}
\centering
\includegraphics[totalheight=1.1in]{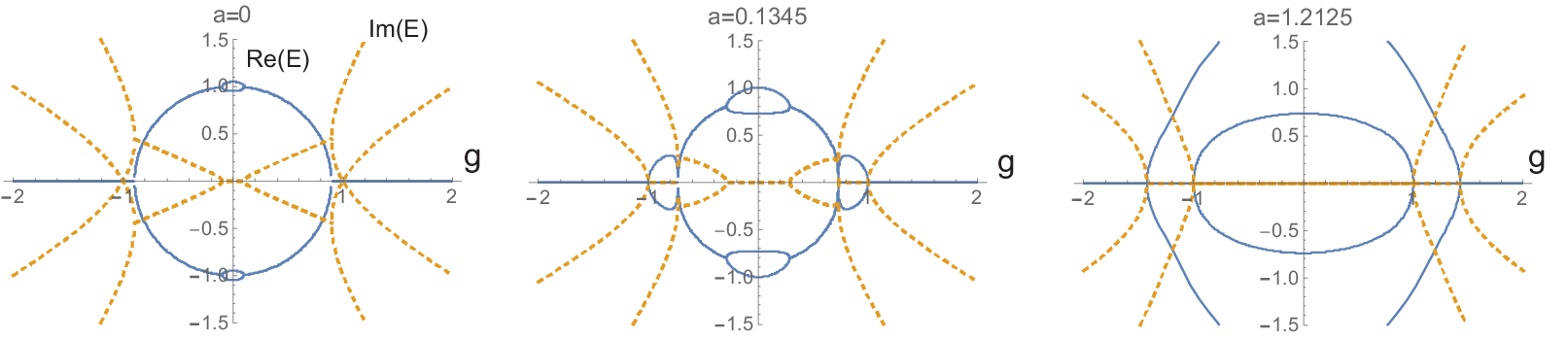}
\caption{Examples of the spectrum versus $g$ for $|v|\ge |w|$. The values $v=1.0$, $w=0.1$ were used. The solid blue curve is the real part of the energy and the dashed yellow curve is the imaginary part. }\label{vgewspecfig}
\end{figure}

Representative examples of the spectrum versus $g$ are shown in Figs. \ref{vlewspecfig} and \ref{vgewspecfig} for several values of $a$. It can be seen that the spectrum is very different for the cases $|v|\le |w|$ and $|v|\ge |w|$, which would correspond to the topologically trivial and nontrivial cases for the standard SSH model. The structure varies considerably as the gradient parameter $a$ is varied, with exceptional points appearing, disappearing, colliding, and merging. In addition, there are bifurcations of the real and imaginary parts at points well removed from the $E=0$ axis. The goal below is to classify this complicated behavior in a relatively simple manner.

First, consider the nonreciprocal case ($g\ne 0$) with no gradient ($a=0$).
Setting $a=0$ in Eq. \ref{H4} the eigenvalues are readily found to be of the form
\begin{equation}E= \pm \sqrt{A\pm B} , \label{E4}\end{equation} where
\begin{eqnarray}A&=& -{3/2} g^2+v^2+{1/2 }w^2 \label{A4} \\
B&=& \sqrt{ A^2 -(g^2-v^2)}. \label{B4} \end{eqnarray}

There are two potential sources of EPs, leading to three opposite-sign pairs of EPs. One possibility occurs when the expression inside the square root of Eq. \ref{E4} passes through zero. This happens when $A^2=B^2$, or equivalently when \begin{equation}g=\pm v. \end{equation}  These EPs always occur at zero energy, and will be referred to as type I points.

The second possibility (type II points) is for the expression inside the square root of Eq. \ref{B4} to pass through zero; this happens when
\begin{equation}g=\pm v\sqrt{{2\over 5} (2v^2+{1\over 2}w^2)} ,\label{gg1}\end{equation} or \begin{equation}g=\pm w.\end{equation} The energies at these points are $E=\pm A$ which can be either zero or nonzero.

Now allow $a\ne 0$. Conceptually, it is expected that the addition of hopping gradients should lead to interesting behavior. This is because, as one moves along the chain, there will be points where the magnitudes of the effective local hopping coefficients will cross each other, with transitions between regions with $|v-ma|>|w-ma|$ and regions with $|v-ma|<|w-ma|$ as the site index $m$ increases. Thus, the gradient would be expected to cause transitions between topological phases in the chain's bulk.

With the gradient turned on, the energies are still of the form of Eq. \ref{E4}, but with $A$ and $B$ now being given by
\begin{eqnarray}A&=& \Big(-{3\over 2} g^2+v^2+{1\over 2 }w^2\Big) +\Big( {5\over 2}a^2-2av-aw\Big) \label{A4a} \\
B&=& \sqrt{ A^2 -(g^2-v^2)^2 +4a(a-v)(g^2-v^2)}. \label{B4a} \end{eqnarray}

It is interesting to note that the energies only depend on the geometric means of the Hamiltonian entries on opposite sides of the diagonals, i.e. opposite-direction hopping amplitudes on the same link. To be more specific, define the geometric means \begin{eqnarray}m_1^2&=& (v+g)(v-g) \\ m_2^2&=& (w+g-a)(w-g-a) \\ m_3^2&=& (v+g-2a)(v-g-2a);\end{eqnarray} then the constants $A$ and $B$ can be written as \begin{eqnarray}A&=&{1\over 2} \left( m_1^2+m_2^2+m_3^2\right) \\ B&=& \sqrt{A^2 +(m_3^2-2m_1^2)m_1^2 }\end{eqnarray}

With $a\ne0$, the type I EPs split into two pairs of points separated by a distance proportional to $a$:  \begin{equation}g=\pm v ,\; \pm (2a-v). \label{typeIIa} \end{equation} The type II points become more complicated, but can still be written in analytic form:
\begin{eqnarray}
g&=& \pm \sqrt{Q\pm 2\sqrt{P}}, \label{gg2}\\
P&=& -69a^2 +(v^2-7a^2)^2 +(w^2+a^2)^2 \\ & & \quad +4a(9a^2v+a^2w-v^3-2avw+v^2w+vw^2-w^3) ,\nonumber \\
Q&=& 12a^2 +2(v-a)^2+3(w-a)^2.
\end{eqnarray} Note the double square root structure of Eq. \ref{gg2} compared to the single square root of Eq. \ref{gg1}; this is what opens the possibility of fourth order exceptional points occurring when $a$ is nonzero.

\begin{figure}
\centering
\includegraphics[totalheight=1.7in]{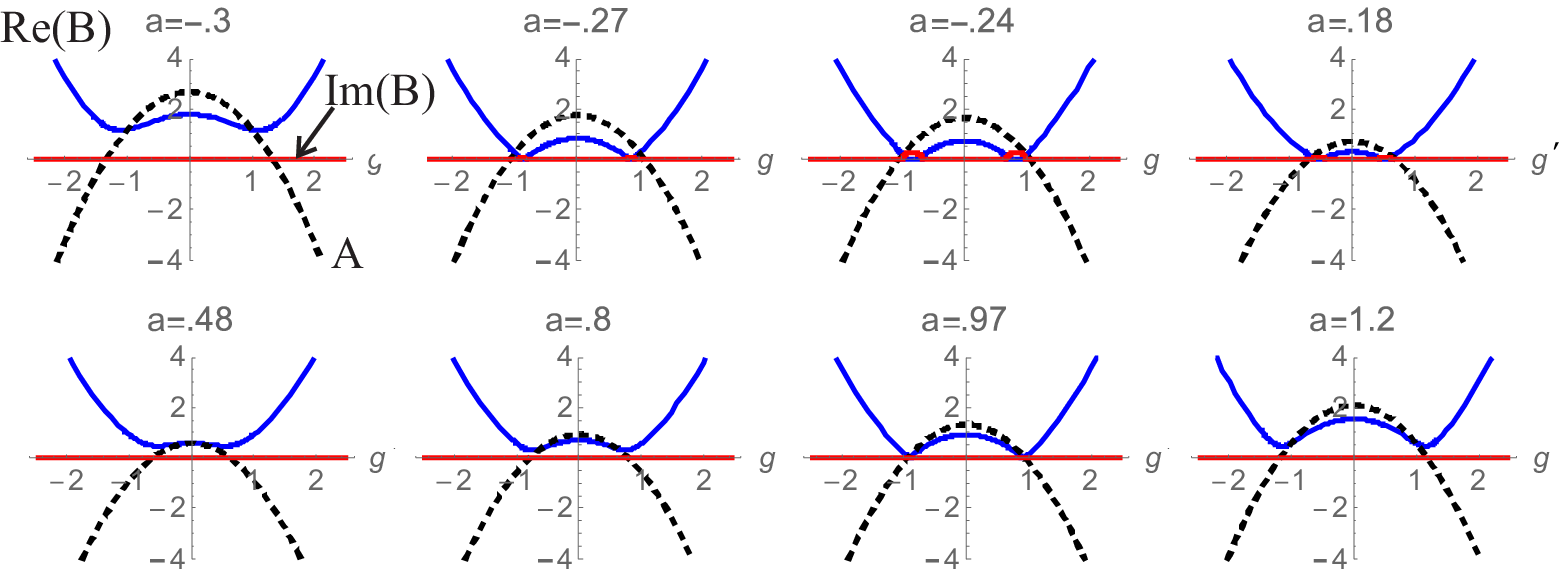}
\caption{$A$ and $B$ versus non-reciprocity parameter $g$ for an example with $v=1.0$, $w=0.1$. The solid blue and red curves and the dashed black curve are, respectively, $Re(B)$, $Im(B)$, and $A$. As the gradient parameter $a$ varies, EPs are created or removed when the $A$ and $B$ lines touch each other or the horizontal axis. A more detailed description of the behavior can be found in the text.}
\label{AB2fig}
\end{figure}

As can be seen from Fig. \ref{AB2fig}, as $a$ is varied the curves of $A$ and $B$ versus $g$ move up and down, while the real part of $B$ can also develop a double-well shape. As a result, the number of intersections of the $A$ and $Re(B)$ curves changes. At these transition points, the imaginary part of $B$ can become nonzero, creating an exceptional point.

%

To study the behavior of the system near the exceptional points more quantitatively, it is necessary to introduce perturbations. So consider perturbations in the gradient, $a\to a+\epsilon$, with real $\epsilon<<a$.
Under such a perturbation, it is found that
\begin{eqnarray}
E& =& \pm \Big\{ A_0 +(5a-2v-w)\epsilon
\pm \sqrt{B_0+4(2a-v)(g^2-v^2)\epsilon}\Big\}^{1/2}\label{Ep1} \\
A&=& A_0+(5a^2-2v-w)\epsilon \label{Ap1}\\
B&=&  \sqrt{B_0^2 +4(2a-v)(g^2-v^2)\epsilon },\label{Bp1}
\end{eqnarray} where $A$ and $B$ have been expanded to first-order in epsilon, and zero subscripts indicate the unperturbed expressions.  The Type II case above can now be split into two distinct cases, so that there are now four cases total (including the trivial case):
\vskip 3pt
\noindent\textbf{Case 0: $B_0\ne 0,\; A_0^2 \ne B_0^2$:}

In this case, expanding the square roots in Eqs. \ref{Ep1} and \ref{Bp1}, the result will be linear in $\epsilon$, so that no EP occurs.

\vskip 3pt
\noindent\textbf{Case I: $B_0\ne 0,\; A_0^2 = B_0^2$:}

If $A_0$ vanishes, the constant ($\epsilon$-independent) term inside the square root of Eq. \ref{Ep1} vanishes, leaving a leading term of order $\sqrt{\epsilon}$. There is therefore a second-order EP at the parameter values given by Eq. \ref{typeIIa}.

\vskip 3pt
\noindent\textbf{Case IIA: $B_0= 0,\; A_0^2 \ne B_0^2$:}

Now, using the binomial approximation, the energy is
\begin{eqnarray} E&=& \pm\sqrt{E_0^2 \pm 4(2a-v)(g^2-v^2)\sqrt{\epsilon}}+{\cal{O}(\epsilon)}\\
&\approx& \pm E_0 \Big[ 1\pm {{4(2a-v)(g^2-v^2)}\over {2E_0^2}}\sqrt{\epsilon}  \Big] ,
\end{eqnarray}
giving a second-order EP at the solutions to $B_0=0$.

\vskip 3pt
\noindent\textbf{Case IIB: $A_0^2 = B_0^2 =0$:}

Now,
\begin{eqnarray} E&=& \pm\sqrt{(5a-2v-w)\epsilon \pm 4(2a-v)(g^2-v^2)\sqrt{\epsilon}} \nonumber \\
&\approx& \pm \sqrt{\pm 1} \sqrt[4]{4(2a-v)(g^2-v^2)}\cdot \sqrt[4]{\epsilon} +{\cal O}(\epsilon^{3/4}),
\end{eqnarray}
giving fourth-order EPs at the solutions to $B_0=0$. Each fourth-order EP is formed in the collision of two second-order EPs.

Case IIB is the overlap of cases I and II. So the situation can be readily summarized as follows: if a point in parameter space falls into either category I or II there is a second-order EP at that point, while if it falls into both classes simultaneously (IIB), it becomes fourth-order.

Case (IIB) occurs when the parameter values are given by either
\begin{eqnarray} a&=&{1\over 7} \Big[ 4v-w\pm\sqrt{9v^2-8vw+8w^2}\Big],\\
g&=&\pm (v-2a),
\end{eqnarray}
or
\begin{eqnarray}a&=&{1\over 5}(2v+w)\pm \sqrt{(2v+w)^2-5(w^2-v^2)} ,\\ g&=& \pm v .
\end{eqnarray}

These cases are summarized in Table~\ref{casetable}.  Note that when $a=0$, case IIB occurs at the topological transition point of the standard SSH model, $|v|=|w|=|g|$.

\renewcommand{\arraystretch}{1.5}
\begin{table}\begin{center}
\begin{tabular}{ |l|c| }
 \hline
Case & EPs \\ \hline\hline
(0) $B_0 \ne 0$, $A_0^2\ne B_0^2$ & No EP \\ \hline
(I) $B_0 \ne 0$, $A_0^2 = B_0^2$ &  2nd-order EPs at $g=\pm v,\; \pm (2a-v)$ \\
& ($a$ arbitrary)\\ \hline
(IIA) $B_0 = 0$, $A_0^2\ne B_0^2$ & 2nd-order EPs at solutions to $B_0=0$ \\ \hline
(IIB) $B_0 = 0$, $A_0^2= B_0^2$ &  4th-order EPs at $g=\pm (2a-v)$,\\
&  $a={1\over 7}\left[ 4v-w\pm \sqrt{9v^2-8vw+8w^2}\right]$\\ \hline
\end{tabular} \caption{Summary of the exceptional points for the length-4 chain, in terms of the non-reciprocity and gradient parameters, $g$ and $a$.}\label{casetable}
\end{center}\end{table}

\begin{figure}
\centering
\includegraphics[totalheight=2.0in]{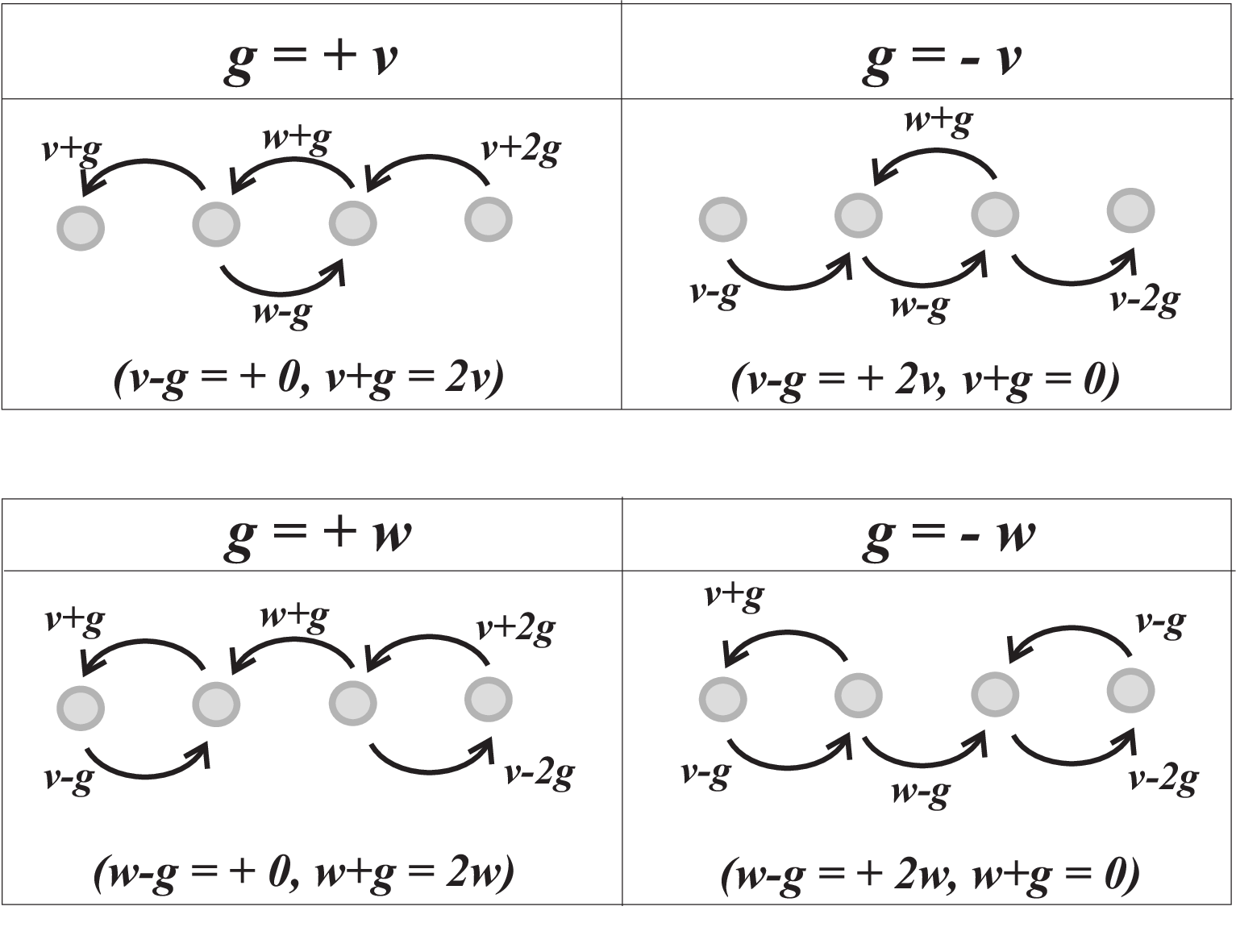}
\caption{Cases for $B_0^2=A_0^2\ne 0.$ In the top row, $g=\pm v$, causing one of the end links to become one-way and leaving all of the amplitude funneled to the site at the left or right end of the chain. Similarly, in the bottom row $g=\pm w$, causing the middle link to become one-way. In the latter case, the amplitude still becomes localized at one end or the other, but oscillating between the last \emph{two} sites. }
\label{onewayfig}
\end{figure}

A coherent picture of these cases can now be constructed. Again consider $a=0$ first. When $B_0^2=A_0^2$ (Case I), we have $g=\pm v$, in which case either the $v+g$ or $v-g$ terms in $H$ vanish. Then the links at the end of the chain become one-way, as in the top row of Figure~\ref{onewayfig}. As a result, the amplitude tends to accumulate at one site at the left or right of the chain.  The parameter-space points where these ``traps'' or attractors occur turn out to be especially interesting when $a$ becomes nonzero. For $a=0$, the real parts of the energies cross the $E=0$ axis, but no EP occurs at the crossing points, since the imaginary parts between the two points remain zero.  But as $a$ starts to move away from zero, these two points split into four: two points remain fixed at $g=\pm v$, while the other pair, at $g=\pm(v-2a)$,  moves away from the first pair. As these two pairs separate, an arc of nonzero imaginary part connects $v$ to $v-2a$ and another such curve connects $-v$ to $-(v-2a)$, so that all four points become EPs. Turning on the gradient term therefore creates four new exceptional points at $g=\pm v, \pm (v-2a).$

Now consider Case II ($B_0=0$). One possibility is that $g=\pm w$, in which case the middle link becomes one-way, as in the bottom row of Figure~\ref{onewayfig}. In this latter case, the amplitude eventually becomes localized in the last two links of the chain at one end or the other. Here, both the real and imaginary parts of the energy bifurcate, breaking degeneracies. One part (real or imaginary) is already zero at the bifurcation point, while the other is nonzero. At these points, two of the eigenvectors vanish.

The other possibility for case II at $a=0$ is when the nonreciprocity parameter becomes $g=\sqrt{{2\over 5} (2v^2+{1\over 2}w^2)}$, which then implies that $E=\pm \sqrt{A_0}=\sqrt{{1\over 5}(w^2-v^2)}$. Assuming that $A_0\ne 0$, these are points at which bifurcations occur at nonzero energies; the real and imaginary parts separately bifurcate, one starting from zero and one starting from a nonzero value. These points are second-order EPs, at which two eigenvectors merge.

If $a=0$, the additional constraint $A_0=0$ cannot be satisfied concurrently with $B_0=0$ (except in the trivial case $v=w=0$), but
allowing nonzero $a$ opens access to points at which Case IIB occurs.  At such points, the $E\ne 0$ bifurcations mentioned in the previous paragraph will now occur at $E=0$, and two pairs of second-order EPs collide, forming fourth-order exceptional points.  These collisions and mergers can be seen clearly, for example, between the second and third images of Figure~\ref{vlewspecfig}, where pairs of pitchfork-shaped imaginary parts move onto the real axis and coincide with each other.
As $a$ is increased further, the merged points separate again; after the separation, they remain on the real axis, forming two second-order EPs.

\begin{figure}
\begin{center}
\subfigure[]{
\includegraphics[height=1.4in]{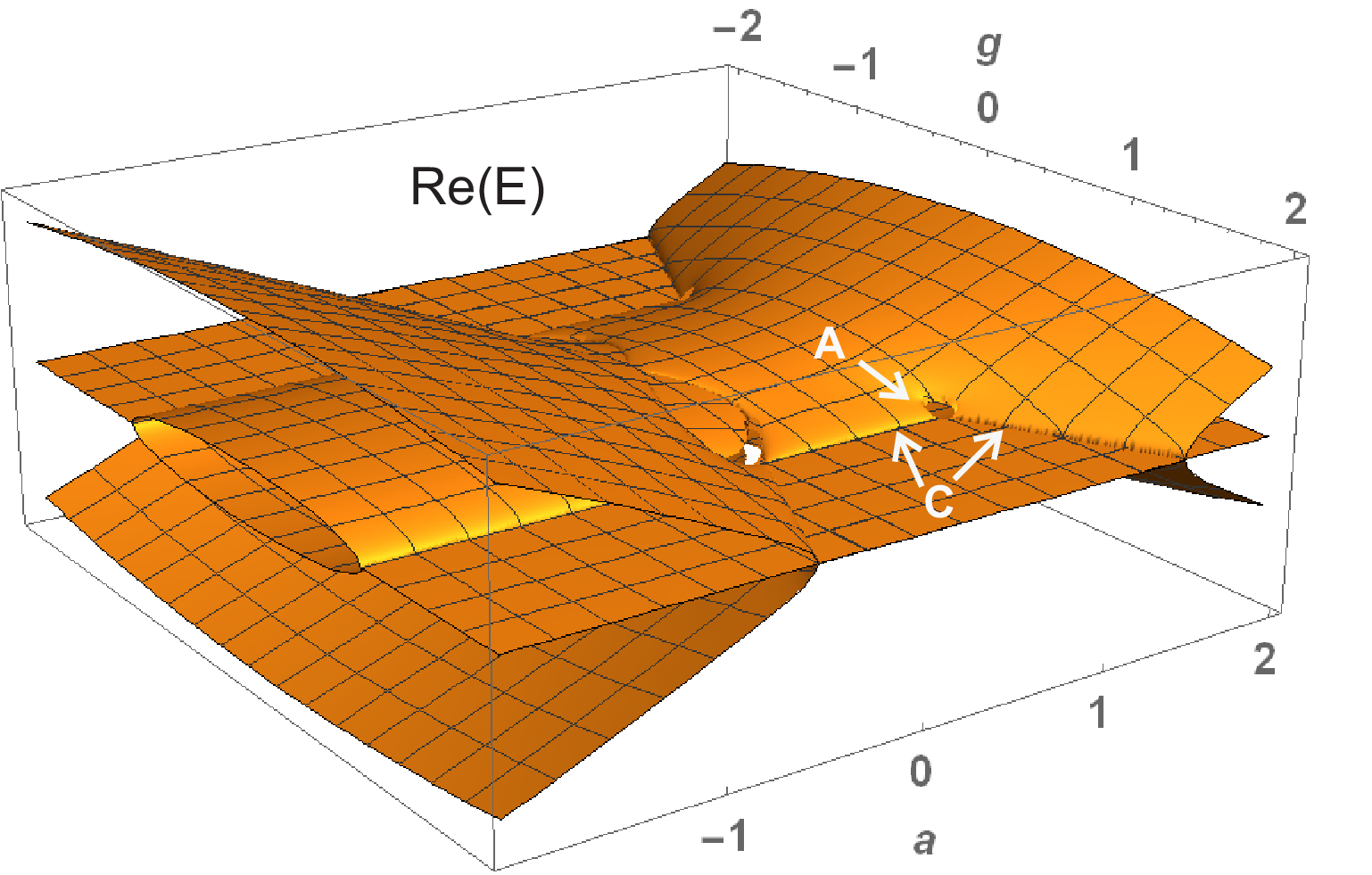}\label{3drefig}}
\subfigure[]{
\includegraphics[height=1.4in]{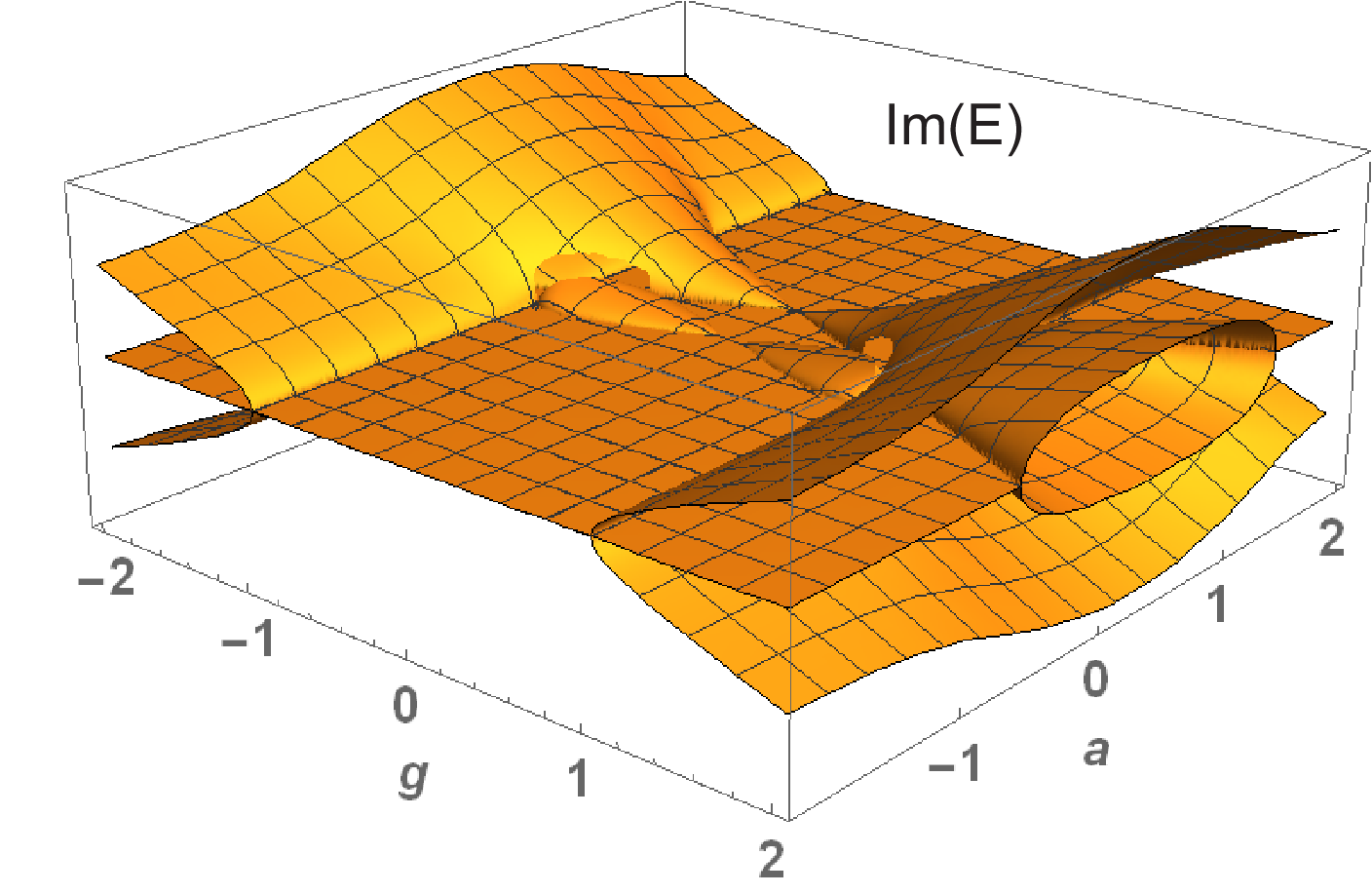}\label{3dimfig}}
\caption{Real (a) and imaginary (b) parts of the energy as function of nonreciprocity $g$ and gradient $a$, for a chain of length $N=4$. The hopping parameters were taken to be $v=1.0$, $w=0.1$ . In (a), the curves marked $C$ are examples of lines of second order EPs. The intersections of two of these at points such as the one marked $A$ lead to fourth order EPs.} \label{3dH4fig}
\end{center}
\end{figure}

The full evolution of the eigenvalues can be seen when plotted as a function of both $g$ and $a$ (Figure~\ref{3dH4fig}). It can be observed that for $N=4$, the real and imaginary parts both consist primarily of three sheets, which split for some parameter ranges into opposite sign subsheets. The bifurcation points form curves of exceptional points in the parameter space; two examples of such curves are marked $C$ in Figure~\ref{3drefig}. The fourth-order EP emerges when two of these curves intersect, for example at the point marked $A$ in the Figure.

\begin{figure}
\begin{center}
\subfigure[]{
\includegraphics[height=1.4in]{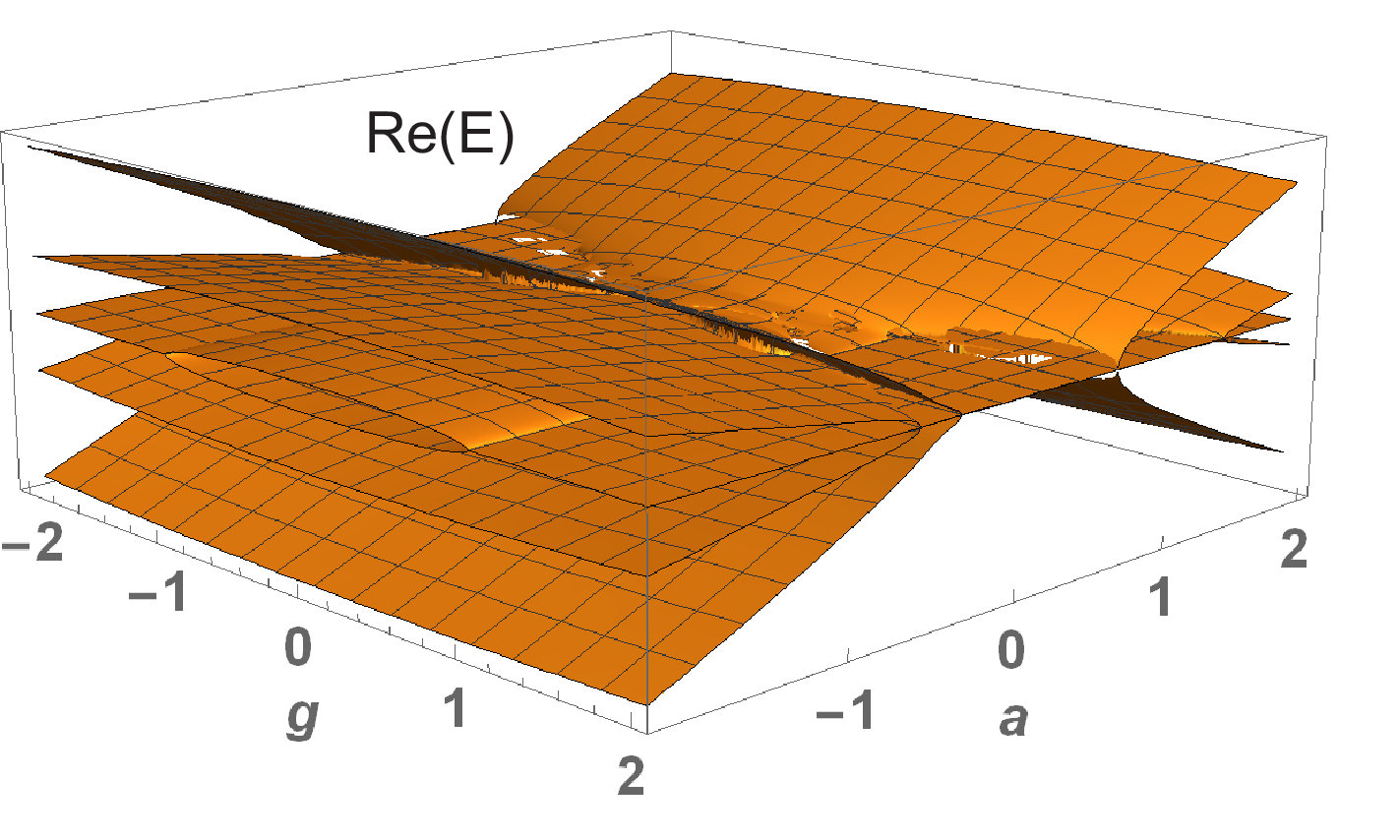}\label{3dHrealfig}}
\subfigure[]{
\includegraphics[height=1.4in]{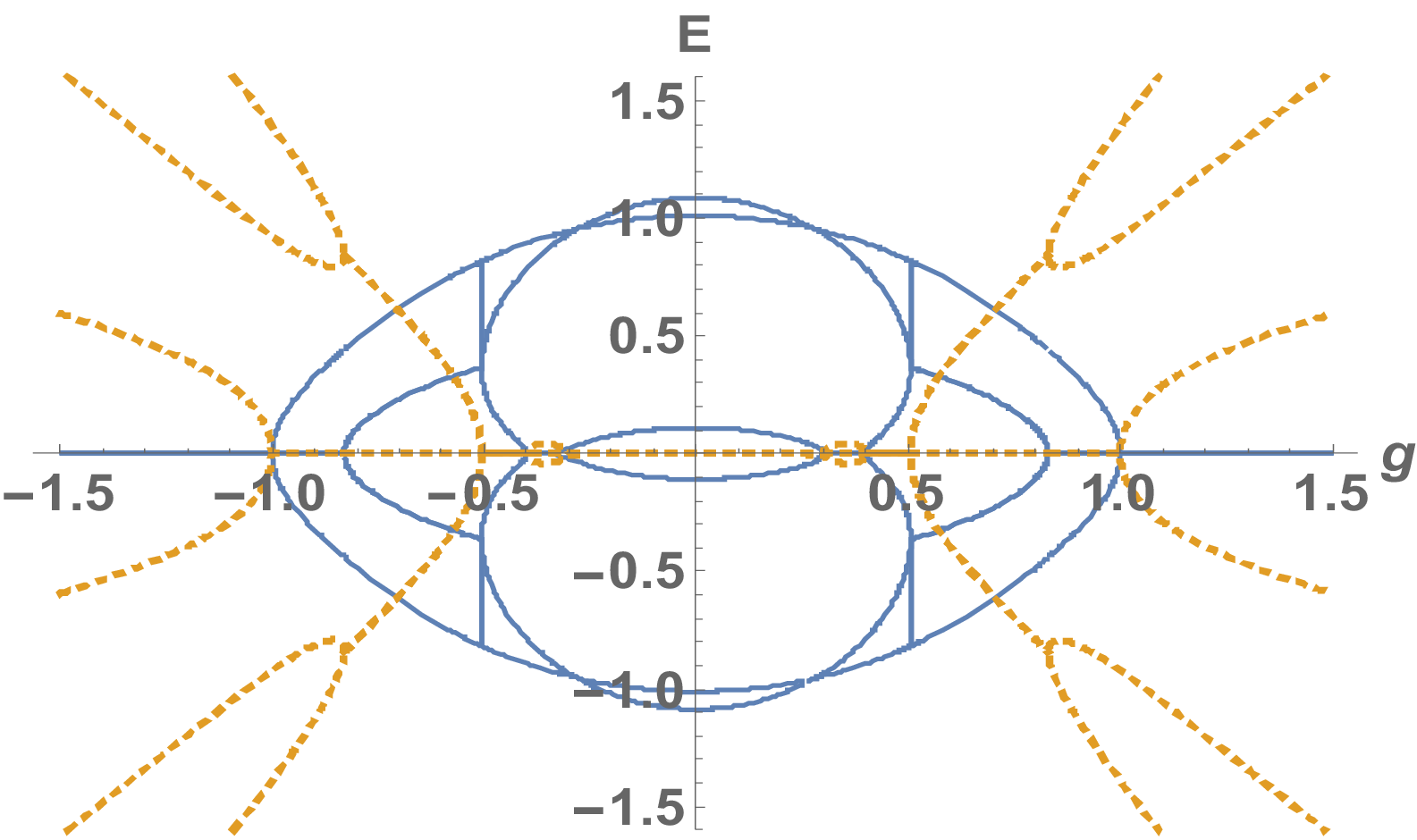}\label{H6sectionfig}}
\caption{(a) Real part of the energy as function of nonreciprocity $g$ and gradient $a$, for a chain of length $N=6$. The hopping parameters were taken to be $v=1.0$, $w=0.1$ . (b) An example of a cross-section of this plot for $a=0.35$. The solid blue curves are the real parts and the dashed yellow curves are imaginary parts.}\label{H6fig}
\end{center}
\end{figure}

The pattern persists for longer length chains, where the number of principle sheets is $N-1$, with each sheet splitting at exceptional point curves. The $N=6$ case is shown as an example in Figure~\ref{H6fig}. While, for the $N=4$ case there are four second-order EPs for $a=0$, which then split or merge as $a$ is varied, in the $N=6$ case there are eight EPs at $a=0$, which then undergo similar evolution as $a$ changes. As the chain length increases, the number of spectral sheets and the number of EPs both increase, with the spectrum rapidly becoming extremely complicated.

\section{Possible experimental implementations}\label{experiment}

Optical and photonic platforms (see Refs. \cite{sav,rap,piec}, among others) and electric circuits \cite{lee} have been implemented that provide physical embodiments of the standard SSH system. Many of these can be altered to implement the model of the previous section, providing a means of easily producing a highly-controllable set of EPs in systems whose behavior can be altered in real time.

\begin{figure}
\centering
\includegraphics[totalheight=1.4in]{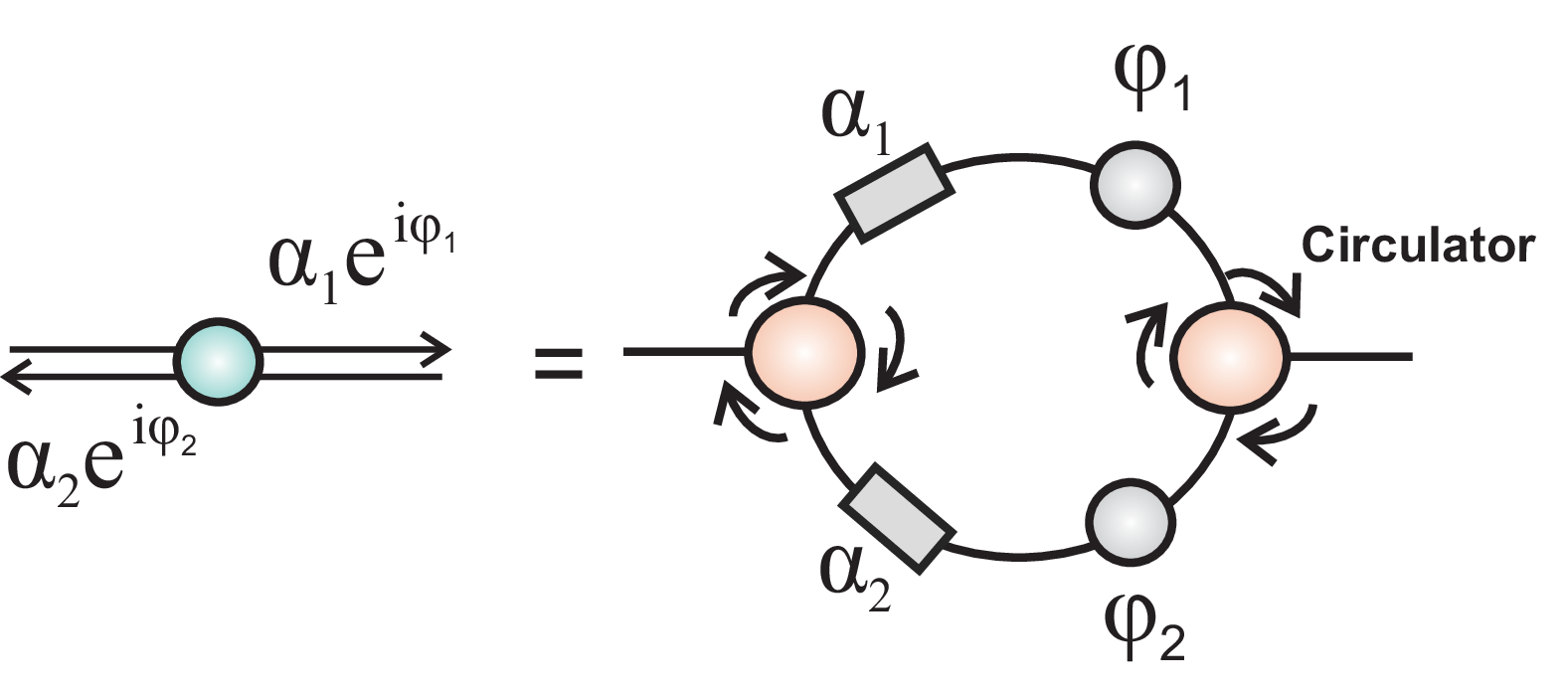}
\caption{Optical implementation of non-reciprocal hopping. The circles on the two sides are optical circulators. The upper and lower lines in the left-hand figure contain loss represented by $\alpha_j$ and a phase shift $\phi_j$. $\alpha_j$ is the fraction of amplitude that remains after the loss. The left-hand figure will be used as shorthand for the arrangement on the right. }\label{Fig2fig}
\end{figure}

For example, this model can be realized with linear optics, augmented by optical circulators.
Non-reciprocal hopping between adjacent lattice sites can be arranged using a pair of circulators, a pair of phase shifters, and some loss. The circulators make use of magnetic fields, and therefore break time reversal symmetry. The hopping amplitudes can be implemented in an arrangement of the form shown in Fig. \ref{Fig2fig},
where the circles on the two sides are optical circulators. The circulators ensure that amplitude can only flow to the right in the upper branch and only toward the left in the lower branch. Denoting the upper and lower branches by labels $1$ and $2$, respectively, the amplitude in each line encounters a phase shifter with phase $\phi_j$ for $j=1,2$. The losses are represented by real, positive parameter $\alpha_j$. The amplitude for transmission to the right is then $\alpha_1e^{i\phi_1}$, with transmission to the left being given by $\alpha_2 e^{i\phi_2}$.
The losses can be implemented by inserting Mach-Zehnder interferometers with one output port used to eject light from the system; by altering the phase in the interferometer the amount of loss through this port can be controlled. The magnitudes of hopping amplitudes can also be easily controlled, again by use of Mach-Zehnder interferometers (or by temperature control if the system is implemented on an integrated chip). So this optical implementation allows \emph{all} parameters to be tuned in real time via experimenter-controlled phase shifts.

\begin{figure}
\begin{center}
\subfigure[]{
\includegraphics[height=.9in]{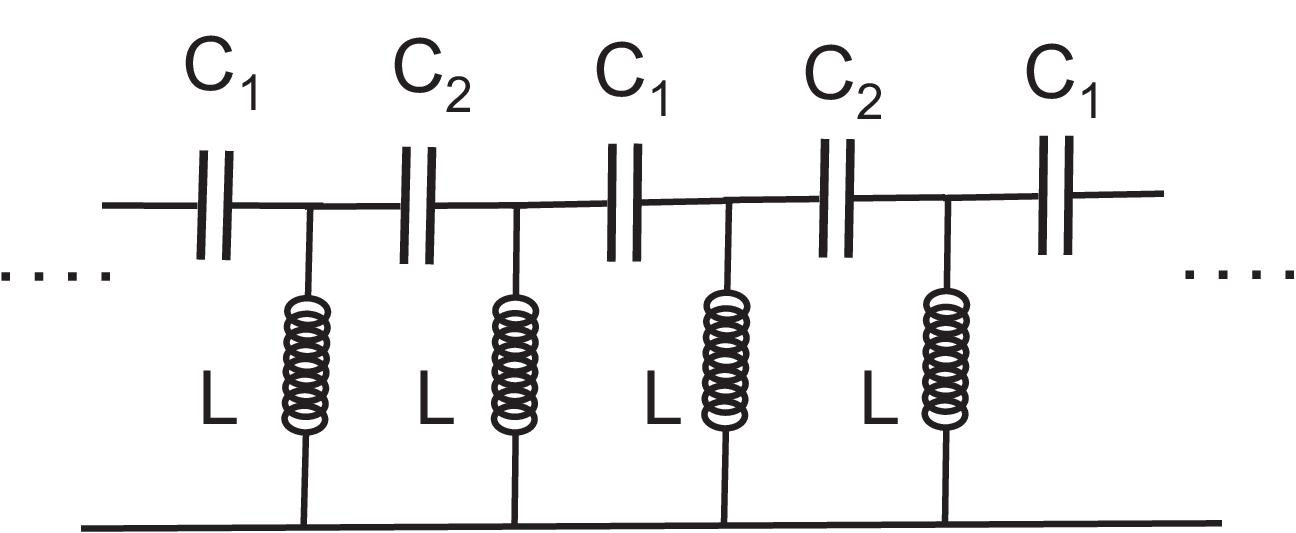}\label{circuitfig}}\quad
\subfigure[]{
\includegraphics[height=.9in]{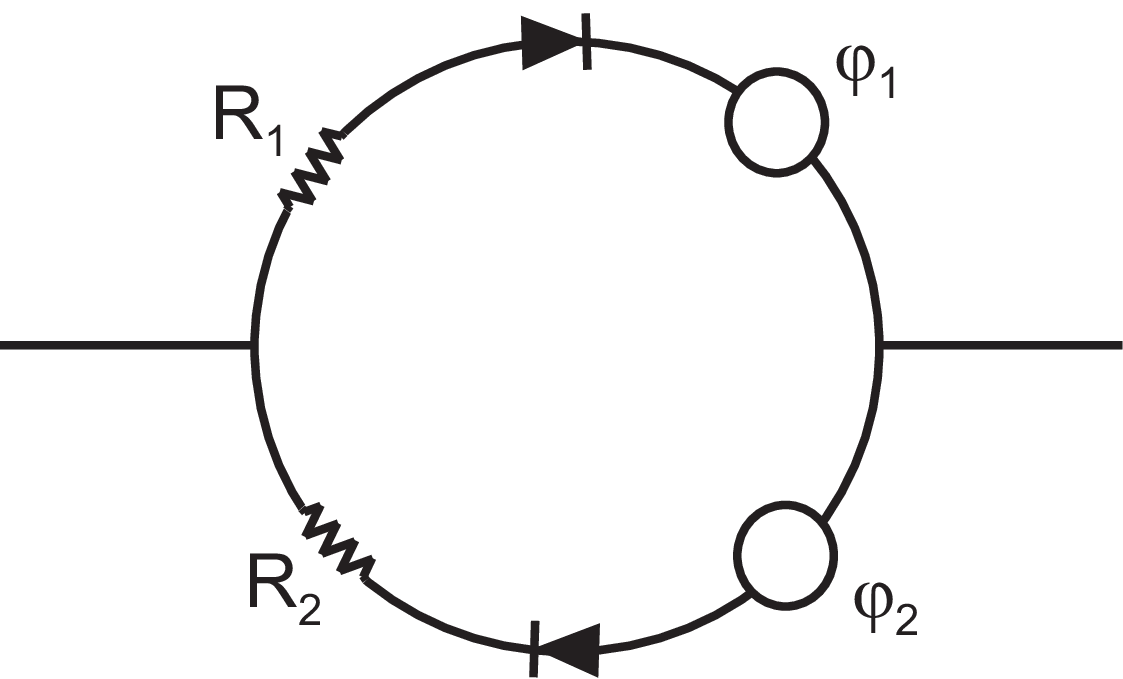}\label{circbfig}}
\caption{(a) An electric circuit implementation of the standard Hermitian SSH model. (b) Electrical analog of the module in Fig. \ref{Fig2fig}. The phase shifts can be implemented by appropriate inductor-capacitor combinations.}
\end{center}
\end{figure}

Electric circuits provide another viable platform. A circuit implementation of the standard SSH model \cite{lee} is shown in Fig. \ref{circuitfig}. Each unit cell consists of pairs of inductors and capacitors, with the capacitance values alternating from one cell to the next. A hopping gradient could easily be introduced into this system by varying the capacitances from one cell to the next. Loss and nonreciprocality are readily introduced by the addition of resistors and diodes (Fig. \ref{circbfig}). The use of variable resistors and capacitors allows the loss and phase shifts to be controlled in real time.

Both optical and electrical implementations can easily be constructed (either on a table top or in integrated form) with low resource cost and allow real time control of all system parameters. This greatly simplifies experimental study of the topological and singularity structures of the system, and allows band gaps to be opened or closed at will. Different experiments may be run on the same apparatus, with only a change of a few resistance and capacitance values, or the change of a few phase shifts in the optical case. In either case, this is achieved with the turn of a knob, rather than a physical alteration of the apparatus.

\begin{figure}[t!]
\centering
\includegraphics[totalheight=1.0in]{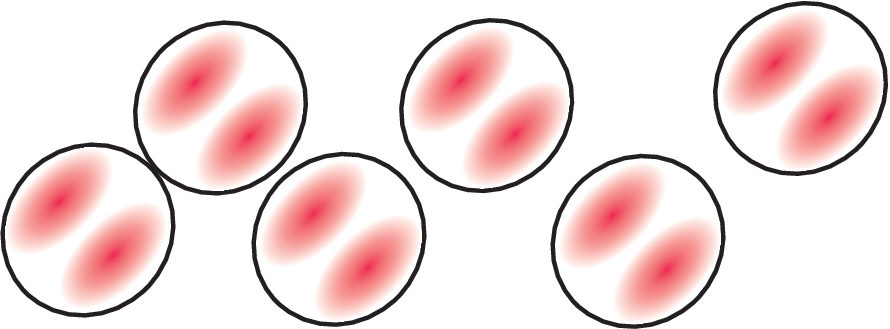}
\caption{Lasing occurs in staggered nanopillar SSH arrays \cite{stjean}. By gradually increasing the spacing between pillars from left to right, a gradient in the hopping amplitudes can be introduced. Heating of the substrate can lead to a hopping amplitude perturbation. The shaded lobes represent the amplitudes of $p$-state polariton excitations. }\label{laserfig}
\end{figure}

It is possible that a nanopillar array (Fig. \ref{laserfig})
such as that used to implement SSH-based topological lasers in Ref. \cite{stjean}, can also be arranged to implement the desired gradient, simply by gradually increasing the spacing between adjacent staggered pillars. Coupled microring cavities could be used in a similar manner. Introducing non-reciprocal transitions is more difficult, but may be possible using recent advances in integrated photonics, such as the integrated isolator developed in Ref. \cite{tian}.

\section{Conclusions}

In the previous sections, a non-Hermitian model has been presented and analyzed which generalizes the SSH model by introducing both nonreciprocal couplings and spatial variation of the coupling constants. The nonreciprocity creates exceptional points, while the introduction of the hopping gradients affects the number and nature of the EPs. By implementing this system in a physical platform that allows the hopping gradients to be controlled, this introduces a means of tailoring the EPs to have desired properties or be located at desired points in parameter space. In particular, if the gradients can be varied in a user-controllable manner, it allows experimental study of new situations such as the braiding of exceptional points around each other in parameter space.

The model studied here is the simplest in a class of models that has a rich and unexplored expanse of variations. By allowing the hopping coefficients to vary in a nonlinear manner, or by allowing the nonreciprocity parameter $g$ to also be spatially-varying, additional control over the spectrum can be introduced, with the potential of uncovering further interesting and useful effects.

\section*{Funding}
This research was supported by the Air Force Office of Scientific Research MURI Award No. FA9550-22-1-0312, DOD/ARL Grant No. W911NF2020127, and the Beckman Young Investigator Award, from the Arnold and Mabel Beckman Foundation.

\section*{Disclosures}
The authors declare no conflicts of interest.

\section*{Data Availability Statement}
All data generated for this paper is available upon reasonable request.

\end{document}